\documentclass[aps,preprint,showpacs,amsfonts,amsmath,amssymb,prb]{revtex4}

\usepackage{graphicx}
\usepackage{dcolumn}
\usepackage{bm}
\usepackage{longtable}

\begin{document}

\title{Absence of long-ranged charge order in $\rm Na_{x}Ca_{2-x}CuO_{2}Cl_{2}$ at x=0.08}

\author{\c{S}erban Smadici}
  \affiliation{National Synchrotron Light Source, Brookhaven National Laboratory, Upton, NY 11973-5000, USA}%
  \affiliation{Frederick Seitz Materials Research Laboratory, University of Illinois, Urbana, IL 61801, USA}%
\author{Peter Abbamonte}
  \affiliation{National Synchrotron Light Source, Brookhaven National Laboratory, Upton, NY 11973-5000, USA}%
  \affiliation{Frederick Seitz Materials Research Laboratory, University of Illinois, Urbana, IL 61801, USA}
\author{Munetaka Taguchi}
  \affiliation{RIKEN/SPring-8, Mikazuki, Hyogo 679-5148, Japan}
\author{Yuhki Kohsaka}
\affiliation{Department of Advanced Materials, University of Tokyo,
Kashiwa, Chiba 277-8562, Japan}
\author{Takao Sasagawa}
\affiliation{Department of Advanced Materials, University of Tokyo,
Kashiwa, Chiba 277-8562, Japan}
\author{Masaki Azuma}
\affiliation{Institute for Chemical Research, Kyoto University,
Kyoto 611-0011, Japan}
\author{Mikio Takano}
\affiliation{Institute for Chemical Research, Kyoto University,
Kyoto 611-0011, Japan}
\author{Hidenori Takagi}
  \affiliation{Department of Advanced Materials, University of Tokyo, Kashiwa, Chiba 277-8562, Japan}
  \affiliation{CREST-JST, Chiba 277-8562, Japan}
  \affiliation{RIKEN, Wako 351-0198, Japan}

\begin{abstract}
A periodic $4a \times 4a$ density of states (DOS) modulation (a
``checkerboard pattern") was observed with STM in $\rm
Na_{x}Ca_{2-x}CuO_2Cl_2$ (NCCOC) [T. Hanaguri \emph{et al.}, Nature
\textbf{430}, 1001 (2004)]. Its periodicity is the same as that of
the ``stripe" charge order observed with neutron scattering in $\rm
La_{1.875}Ba_{0.125}CuO_4$ (LBCO) [J. M. Tranquada \emph{et al.},
Nature \textbf{429}, 534 (2004)] and $\rm
La_{1.48}Nd_{0.4}Sr_{0.12}CuO_4$ (LNSCO) [J. M. Tranquada \emph{et
al.}, Phys. Rev. B \textbf{54}, 7489 (1996)]. An obvious question is
whether the ``stripes" are actually ``checkers". Unfortunately,
because NCCOC samples are small and LBCO samples do not cleave,
neutron and STM measurements cannot be carried out on the same
system.  To determine the relationship between stripes and checkers
we used resonant soft x-ray scattering (RSXS), previously applied to
LBCO [P. Abbamonte \emph{et al.}, Nature Physics \textbf{1}, 155
(2005)], to study single crystals of NCCOC.  No evidence was seen
for a $4a \times 4a$ DOS modulation, indicating that the
checkerboard effect is not directly related to the stripe modulation
in LBCO. Our measurements suggest either glassy electronic behavior
or the existence of a surface-nucleated phase transition in NCCOC
[S. E. Brown \emph{et al.}, Phys. Rev. B \textbf{71}, 224512
(2005)].

\end{abstract}


\maketitle

$\rm Na_{x}Ca_{2-x}CuO_{2}Cl_{2}$ (NCCOC) is a high temperature
superconductor with a crystal structure similar to that of $\rm
La_{2-x}Sr_{x}CuO_4$ (LSCO), however with the LaO layers replaced by
CaCl layers. In particular, an apical chlorine atom in NCCOC is
substituted for the apical oxygen in LSCO. The discovery of
superconductivity in this compound was the original proof that
high-$\rm T_c$ superconductivity can occur in the absence of an
apical breathing phonon.~\cite{HKT1996} Because of the absence of
apical oxygens the coupling between $\rm CuO_2$ planes in NCCOC is
decidedly weaker than in LSCO, making NCCOC electronically more
two-dimensional and cleavable.~\cite{KAYS2002} This latter trait has
facilitated studies of NCCOC with optics and transport
probes~\cite{WKKS2004}, angle-resolved
photoemission~\cite{RSKS2002}, and scanning tunneling spectroscopy
(STS)~\cite{HLKL2004,KISH2004}.

A recent STS study of cleaved NCCOC has shown evidence for a
``checkerboard" electronic superlattice with a period of $4 \times
4$ unit cells.~\cite{HLKL2004,KISH2004} The periodicity of this
pattern is very close to that of the charge superlattice observed
with neutron scattering in $\rm La_{1.48}Nd_{0.4}Sr_{0.12}CuO_4$
(LNSCO)~\cite{TAIN1996} and $\rm La_{1.875}Ba_{0.125}CuO_4$
(LBCO)~\cite{TWPG2004}, which is frequently cited as evidence for
charged stripes. Interestingly, NCCOC is by nature tetragonal at all
temperatures and the ``stripe" charge order only forms when LNSCO or
LBCO are in the low temperature tetragonal phase. Because of the
similarity in crystal structures and superlattice periods, these STS
measurements have raised the question of whether NCCOC, LNSCO and
LBCO contain the same phase and, in fact, are all checkerboards
rather than stripes. Unfortunately, because LNSCO and LBCO do not
cleave, and single crystals of NCCOC are extremely small (typically
$\rm 0.5\times0.5\times0.1 mm^3$), it is not possible to do both STS
and neutron scattering on the same system to determine if these
effects are related.

We recently reported a study of the charge order in LBCO with
resonant soft x-ray scattering (RSXS) at the $\rm O~K$
edge.~\cite{PRSG2005} While RSXS cannot easily discriminate between
stripe and checkerboard order, we were able to determine that the
charge order in LBCO is mainly electronic and similar in amplitude
to that claimed for NCCOC.~\cite{NOTE1} RSXS can be performed on
small samples as well as materials that do not cleave. To determine
if the checkerboards in NCCOC are related to the charge order in
LBCO, we have used soft x-ray absorption spectroscopy (XAS) and RSXS
to characterize the electronic structure of NCCOC.

\begin{figure}
\rotatebox{270}{\includegraphics[scale=0.6]{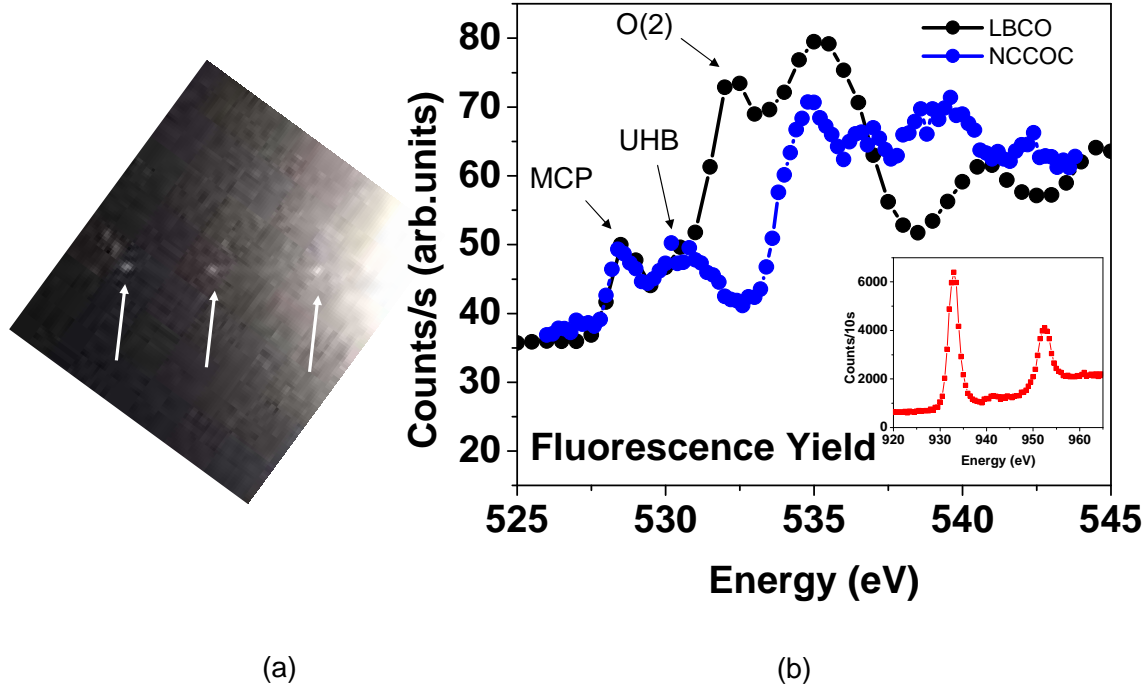}}
\caption{\label{fig:Figure1}(Color online) (a). Detail of Laue image
of a NCCOC crystal obtained with a rotating anode source. The
absence of heavy elements in this compound is responsible for the
reduced contrast. The diffraction peaks shown by the arrows are
aligned along the a-b axes. (b). Absorption spectrum in NCCOC and
LBCO at the OK edge with two pre-peaks at $\rm 528.5~eV$ and $\rm
530.75~eV$. The two pre-peaks are the mobile carrier peak (MCP) and
the upper Hubbard band (UHB). The spectra have been aligned at the
background level. The inset shows the absorption at the $\rm CuL_3$
and $\rm CuL_2$ edges in NCCOC.}
\end{figure}

The NCCOC crystals were grown following preparation conditions
described previously.~\cite{KAYS2002} We investigated two batches,
grown at $\rm 3~GPa$ and $\rm 4~GPa$, with a $\rm x=0.08$ doping at
which charge order was previously reported~\cite{HLKL2004}. NCCOC is
extremely hygroscopic so sample mounting and handling was done in a
nitrogen-filled glove box with $\rm <10~ppm$ $\rm O_2$. The crystals
were oriented {\it ex situ} on a Laue diffractometer prior to x-ray
measurements. To minimize exposure to air the crystals were covered
with a thin polycarbonate foil and the exposure time limited to the
minimum necessary for an unambiguous identification of the $a$ and
$b$ axes. An example of a Laue image is shown in
Fig.~\ref{fig:Figure1}. Prolonged exposure to air causes these
points to broaden into concentric arcs along a direction parallel to
the arrows; the absence of arcs in Fig.~\ref{fig:Figure1} is
indicative of good sample quality.

XAS and RSXS measurements were carried out on the X1B soft x-ray
undulator beam line at the National Synchrotron Light Source with a
ten-axis, ultrahigh vacuum-compatible diffractometer. The focus size
was $\rm 0.5~mm\times1~mm$ allowing investigation of moderately
small samples. Measurements were typically done in a vacuum of $\rm
5\times10^{-9}~mbar$. XAS measurements were made at room temperature
and in fluorescence yield mode to assure bulk sensitivity. The
checkerboard pattern has been observed in STM images up to $\rm
T=30~K$; therefore all RSXS measurements were done at $\rm T=20~K$,
a temperature at which the charge order is stable. In the geometry
used the probe depth was approximately $\rm 1000~\AA$ for XAS and
RSXS measurements.

\begin{figure}
\rotatebox{270}{\includegraphics[scale=0.6]{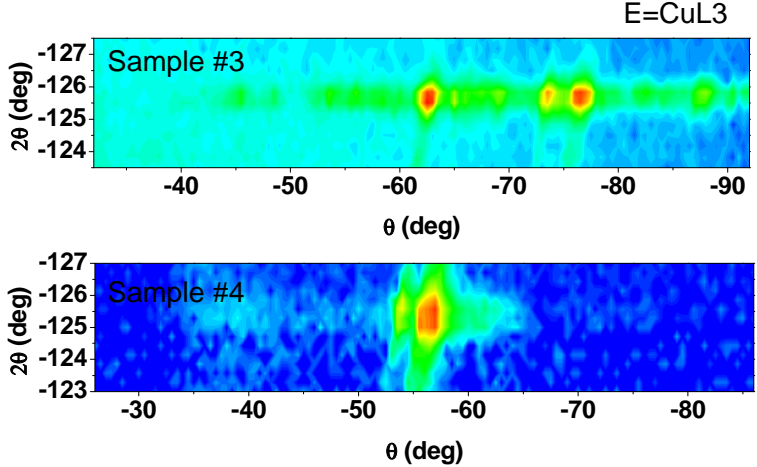}}
\caption{\label{fig:Figure2} (Color online) Scans in the
$\theta-2\theta$ plane at $\rm E=933~eV$ showing the (002) Bragg
peak for a bad (top) and good (bottom) cleave (logarithmic scale).
In the top panel diffraction from many misaligned crystallites is
visible on the high background. In both cases the spots are
elongated in the L direction. The samples are from the same batch.}
\end{figure}

Prior to x-ray measurements the crystals were cleaved at room
temperature in a $\rm 10^{-6}~mbar$ vacuum and immediately
transferred to the UHV chamber. Cleaving sometimes resulted in poor
diffraction maps of the (002) Bragg reflection, indicating a highly
corrugated near-surface region. Examples of good and bad cleaves are
shown in Fig.~\ref{fig:Figure2}. For good cleaves, however, the
angular width of the (002) was resolution-limited. We denote
reciprocal space in this article by Miller indices $\rm (H,K,L)$
which indicate a momentum transfer $\rm {\bf Q} = (2\pi H/a, 2\pi
K/a, 2\pi L/c)$ with $\rm a=3.85~\AA$ and $\rm c=15.1~\AA$.

\begin{figure}
\rotatebox{270}{\includegraphics[scale=0.6]{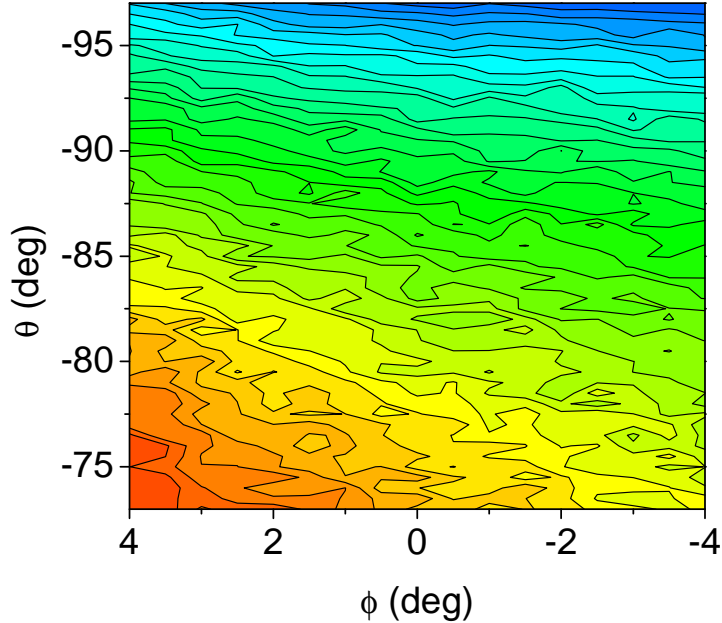}}
\caption{\label{fig:Figure3} (Color online) A two-dimensional scan
at $\rm E=933~eV$ photon energy. This region of reciprocal space is
centered at (0.25, 0, 1.5) and ranges from $\rm H\sim0.15$ to $\rm
H\sim0.35$. The sloping background is due to the variation of sample
absorption with $\phi$ and $\theta$.}
\end{figure}

Absorption spectra of NCCOC and LBCO with $\rm x=1/8$ (from Ref.~9)
are shown in Fig.~\ref{fig:Figure1}. The NCCOC spectra are similar
to previous studies of the closely related insulator $\rm
Sr_2CuO_2Cl_2$~\cite{HNK1998}, however with a pronounced mobile
carrier peak (MCP) band and decreased upper Hubbard band (UHB). This
indicates that spectral weight in the NCCOC system is transferred
with doping from the UHB peak to the MCP peak, as in other
cuprates~\cite{CSMH1991}. It also indicates that our cleaved NCCOC
surface is good within the probe depth of about $\rm 1000~\AA$.

When compared to LBCO, the most striking feature of NCCOC absorption
spectrum is the absence of the peak labeled $\rm O(2)$ in
Fig.~\ref{fig:Figure1}. The tight-binding model~\cite{DPP1989,
M1990,HNK1998} is an useful framework to analyze these differences.
This O(2) peak is absent in $\rm Sr_{14}Cu_{24}O_{41}$
(SCO)~\cite{ABR2004}, which also does not have apical oxygens,
implying that the peak is due to hybridization of apical oxygen
states. In addition, this peak was shown to be visible only for
photon polarizations ${\bf E}\perp\hat{c}$ in $\rm
La_{2-x}Sr_xCuO_4$~\cite{CTKK1992}, which is consistent with it
arising from apical oxygen levels polarized in the $a-b$ plane. In
the tight-binding model apical oxygen $\rm p_{x}$ states hybridize
with a Cu d state or with La d states with an interaction energy of
$\rm 0.3-0.4~eV$ and $\rm 2.6~eV$ in LBCO,
respectively.~\cite{DPP1989,M1990} In the first case, for $\rm
Cu~d-O(2)~p$ hybridization, the energies of the mixed states are
below the Fermi level. In contrast, because the energy of
non-interacting La d levels are high above the Fermi energy, the
energy of a mixed $\rm La~d-O(2)~p$ state is $\rm \sim7-8~eV$ above
the Fermi level, which is close to the measured value of the O(2)
peak. We conclude that the most likely origin of the peak at $\rm
532.4~eV$ in $\rm La_{2-x}Ba_{x}CuO_4$ is La-O(2) hybridization and
its absence is a signature of the reduced dimensionality of NCCOC.

To study the charge order in NCCOC we tuned the x-ray energy to
either the $\rm Cu L_{3/2}$ edge or $\rm O~K$ mobile carrier
peak~\cite{CSMH1991,AVRS2002} and searched for superlattice
reflections. Broad mesh scans in the $\phi$ and $\theta$ sample
angles were performed around selected values of ${\bf Q}$ to allow
for possible misalignment in the Laue images (see
Fig.~\ref{fig:Figure3}). The in-plane components of {\bf Q} were
determined from STS observations~\cite{HLKL2004}, i.e. $\rm H=1/4$
and $\rm K=0$. However, these measurements are two dimensional so do
not suggest a particular value of $\rm L$. In hard x-ray and neutron
scattering studies of LNSCO~\cite{ZVNI1998} and LBCO~\cite{T2005},
however, the charge order peaks are broad in $\rm L$, following a
$\rm sin^{2}(\pi L)$ dependence. Based on these effects, three
values for L were investigated: $\rm L=0.75$, $\rm L=1.5$ and $\rm
L=2$. For geometric reasons the first was done with the photon
energy tuned to the mobile carrier peak~\cite{AVRS2002} below the
$\rm O K$ edge (528.6 eV) and the latter two at the peak of the $\rm
CuL_{3/2}$ edge (933 eV).

The reciprocal space region around (H, K, L)=(0.25, 0, 1.5),
measured at the $\rm CuL_{3/2}$ edge, is shown in
Fig.~\ref{fig:Figure3}. No peak of the type seen in Ref.~9 was
visible in this or several other samples studied.  Negative results
were also obtained around (0.25, 0, 2) and (0.25, 0, 0.75).
Evidently the checkerboard in NCCOC, in contrast to the charge order
in LBCO, is below our sensitivity limit.

There are several possibilities for why this might be. First, while
difficult to judge from the data in Ref.~5, it is possible that the
electronic checkerboard ordering is glassy despite a well-ordered
crystal structure, i.e. the charge amplitude is large but its
correlation length is short. The signature of glassy ordering in
coherent scattering is a broad peak centered at the ordering wave
vector whose integrated intensity is substantial but whose peak
count rate may be low. If the charge ordering is glassy the
scattering is present but below our fluorescence background, i.e. at
least 50 times weaker than the charge scattering from LBCO. This
scenario allows us to place an upper bound on the quantity $\rm
A\cdot\xi^2$, where $\rm A$ is the charge amplitude and $\xi$ is the
in-plane correlation length. Specifically, assuming the charge order
in NCCOC and LBCO have the same c-axis correlation lengths, it must
be that $\rm
I_{NCCOC}/I_{LBCO}=(A_{NCCOC}\cdot\xi_{NCCOC}^2)/(A_{LBCO}\cdot\xi_{LBCO}^2)<
1/50$. For $\rm A_{LBCO}=0.5$ and $\rm \xi_{LBCO}=480~\AA$ we arrive
at $\rm A_{NCCOC}\cdot\xi_{NCCOC}^2<2.3\times10^{3}~hole\cdot\AA^2$.

Another possibility is that the $4\times4$ structure exists only at
the surface. In a recent mean field analysis, Brown \emph{et.
al.}~\cite{BFK2005}, motivated by the lack of a signature of charge
order in the transport properties of NCCOC~\cite{WKKS2004}, showed
that a commensurate charge density wave can be enhanced at the
surface by poor screening or the presence of soft surface phonon
modes. The authors argue that the commensurate checkerboard pattern
in NCCOC is located in the near-surface region only and is an
example of a surface ``extraordinary" phase transition which
precedes a bulk phase transition in NCCOC. The RSXS signal from such
a surface effect would be weaker than the bulk signal by a factor of
$\sim10^3$. Regardless of the possible explanation for the absence
of a superlattice reflection, we conclude that there is no bulk
static long-range charge order in NCCOC.

\begin{acknowledgements}

The authors acknowledge helpful discussions with Tonica Valla,
Eduardo Fradkin, and J. C. S\'eamus Davis. This work was supported
by US Department of Energy.

\end{acknowledgements}


\begin{thebibliography}{}
\bibitem{NOTE1}
The NCCOC superlattice is visible in $\rm dI/dV$ spectra within $\rm
\pm0.1~eV$ bias voltage, indicating a significant hole modulation
with amplitude $\rm N\times \Delta E\times
DOS\sim4~unitcells\times0.1~eV\times1~state/(unitcell\times eV)\sim
0.4~holes$.
\end{thebibliography}
\end{document}